\documentclass[aps,prd,twocolumn,showpacs,preprintnumbers,amsmath,amssymb]{revtex4}
\usepackage{graphicx}
\usepackage{dcolumn}
\usepackage{bm}
\usepackage{amssymb,amsmath}
\usepackage{latexsym}

\allowdisplaybreaks

\begin{document}

\title{Double Michelson/Fabry-Perot interferometer for laser- and displacement-noise-free gravitational-wave detection}
\author{Sergey P. Tarabrin and Sergey P. Vyatchanin}
\affiliation{Faculty of Physics, Moscow State University, Moscow, 119991, Russia} \email{tarabrin@phys.msu.ru}
\date{\today}

\begin{abstract}
In this paper we demonstrate that a double Michelson interferometer with Fabry-Perot cavities in its arms is able to perform laser- and displacement-noise-free gravitational-wave (GW) detection if certain model assumptions are met. Assuming the input mirrors of a single Michelson/Fabry-Perot interferometer can be rigidly attached to beamsplitter on a central platform one can manipulate with interferometer's response signals in a way to cancel laser noise and displacement noise of all test masses except the cental platform. A pair of symmetrically positioned Michelson/Fabry-Perot interferometers with common central platform can be made insusceptible to the later then, thus allowing complete laser- and displacement-noise-free interferometry (DFI). It is demonstrated that the DFI response to GWs of the proposed interferometer is proportional to $f^2_{\textrm{gw}}/\gamma$, where $\gamma$ is the cavity half-bandwidth, that is the strongest DFI response allowed by general relativity.
\end{abstract}
\preprint{LIGO-P080110}

\maketitle

\section{Introduction}\label{sec_intro}
It is widely known that the first-generation laser interferometric GW detectors suffer from the great amount of noises of various nature. The major limiting factors at low and middle frequencies are seismic noise and thermal noise which can be referred to the class of displacement noise of the test masses. At high frequencies photon shot noise is dominant. In the standard quantum limited (SQL) \cite{1992_quant_meas} second-generation detectors, being under preparation, the cause of SQL is the fluctuating force of radiation pressure in the laser beam (back-action noise) pushing the interferometer mirrors in a random manner. Thus standard quantum limitation also aries due to displacement noise.

Each method of suppressing or eliminating of displacement noise proposed up-to-date is only suited for control of only one kind of noise. For instance, active antiseismic isolation will definitely suppress seismic noise but is helpless against thermal noise or quantum radiation pressure. From the other hand, quantum-non-demolition (QND) schemes of measurements \cite{1981_squeezed_light,1982_squeezed_light,1996_QND_toys_tools} are able canceling back-action noise but are certainly not suited for dealing with seismic or thermal noise.

However, recently there has been proposed a revolutionary new method of displacement noise cancelation which simultaneously eliminates the information about all external fluctuating forces but leaves certain amount of information about the gravitational waves. The major idea is to construct such an interferometer that would respond differently to the motion of the test masses and the GWs. Then the proper linear combination of the interferometer responses will cancel the fluctuations of the test masses leaving a non-vanishing information about the GWs. One may find at least two different methods proposed up-to-date.

The first one, described in a series of papers by S. Kawamura \textit{et al.} \cite{2004_DNF_GW_detection,2006_DTNF_GW_detection,2006_interferometers_DNF_GW_detection}, bears on the distributed nature of the GWs. This can be best explained from the viewpoint of a local observer (or the local Lorentz gauge). In such a reference frame interaction of the GW with a laser interferometer adds up to two effects \cite{2005_local_observ,2007_GW_FP_LL}. The first one is the motion of the test masses in the GW tidal force-field. In this aspect GWs are indistinguishable from any non-GW forces since both are sensed by the light wave only at the moments of reflection from the test masses. We will refer to this as the \textit{localized} nature of the forces acting on the test masses. If the linear scale $L$ of a GW detector is much smaller than the gravitational wavelength $\lambda_{\textrm{gw}}$ (the so-called long-wave approximation) then the effect of the GW force-field is of the order of $h(L/\lambda_{\textrm{gw}})^0$, where $h$ is the absolute value of the GW amplitude. Relative motion of the test masses, separated by a distance $L$, in any force field cannot be sensed by one of them faster than $L/c$, thus resulting in the rise of terms of the order of $O[h(L/\lambda_{\textrm{gw}})^1]$ describing time delays which take the light wave to travel between the masses. Second, GW directly couples to the light wave effectively changing the coordinate velocity of light, thus manifesting itself as an effective refraction index. Light wave traveling in such a (boundless) medium acquires the information about the GW in its phase gradually, therefore it is a \textit{distributed} effect having the $O[h(L/\lambda_{\textrm{gw}})^2]$ order in long-wave approximation. Ultimately, from the viewpoint of local observer displacement-noise-free interferometry implies the cancelation of localized effects (GW and non-GW forces) leaving a non-vanishing information about the distributed effect (the direct coupling of the GW to light).

It was pointed in Refs. \cite{2006_DTNF_GW_detection,2006_interferometers_DNF_GW_detection} that in order the GW detector to be a truly displacement-noise-free interferometer it should be also free from optical laser noise since the latter is indistinguishable from laser displacement noise. Their sum is usually called laser phase noise. Cancelation of laser phase noise in interferometric experiments is usually achieved by implementing the differential (balanced) schemes of measurements: in conventional interferometers (such as LIGO) it is the Michelson topology and in DFIs proposed in Ref. \cite{2006_interferometers_DNF_GW_detection} it is the Mach-Zehnder (MZ) topology.

Although DFI detectors that bear on distributed nature of GWs allow complete elimination of displacement noise, the ``payment'' for such a gain is the significant weakening of the GW response at low frequencies. This fact directly follows from the mechanism of noise cancelation: together with displacement noise we also cancel GW terms of the $h(L/\lambda_{\textrm{gw}})^0$ and $h(L/\lambda_{\textrm{gw}})^1$ orders, leaving only the $O[h(L/\lambda_{\textrm{gw}})^2]$-order terms. In addition, as demonstrated in Ref. \cite{2006_interferometers_DNF_GW_detection}, only 3D (space-based) configurations of DFIs allow the 2nd-order response, while the response of 2D (ground-based) detectors \cite{2006_DTNF_GW_detection} is even weaker --- it is proportional to $h(L/\lambda_{\textrm{gw}})^3$. For the ground-based detectors with $L\sim 10^3$ m and $\lambda_{\textrm{gw}}\sim 10^6$ m such a dramatic decrease of susceptibility to GWs ($\sim 10^{-9}$) levels all the advantages of DFI in comparison with traditional detectors dominated by displacement noise.

The second type of DFI detectors proposed in Ref. \cite{2008_DFI_FP_toy_model}, however, is free from this drawback since the mechanism of noise cancelation does not utilize the notion of GW distributed nature. The basic element of the proposed toy model is a single Fabry-Perot (FP) cavity pumped through both its mirrors with orthogonally polarized light waves. Consider first the pump through one of the mirrors. Two responses will correspond to the input wave --- the reflected wave and the transmitted one. The latter is proportional to the distance between the mirrors which may vary in time under the influence of displacement noise and the GW. The reflected wave, however, is somewhat different: besides the term proportional to the variable length of the cavity it also includes the component due to the prompt reflection of the incident wave from the input mirror. Thus, reflected and transmitted waves carry slightly different information about the GW and the motion of the test masses. Combining them properly one is able to exclude displacement noise of one of the cavity mirrors. In a similar way one can deal with the second pump through another mirror. Ultimately, the proper linear combination of all four responses allows elimination of displacement noise of both mirrors leaving a non-vanishing GW signal proportional to $h(L/\lambda_{\textrm{gw}})^0$ in long-wave approximation. Unfortunately, since the mechanism of noise cancelation bears on the effect of prompt reflection which is nonresonant, the obtained DFI response is not amplified with the usual resonant factor associated with the accumulation of low-frequency signal inside the cavity.

It may seem that the elimination of mirrors displacement noise at the $h(L/\lambda_{\textrm{gw}})^0$-order contradicts the force-like behavior of the GWs at low frequencies. However, it has been demonstrated in Ref. \cite{2008_DFI_FP_toy_model} that the loss of the resonant gain results in the sensitivity limitation by displacement noises of auxiliary optical elements such as lasers and detectors. In conventional non-DFI detectors these noises are negligible since they are finesse times smaller as compared to displacement noise of the mirrors. In a double-pumped FP cavity these residual fluctuations mean that the GW-induced displacements cannot be measured absolutely but only with respect to some reference test masses, in full agreement with the relativity principle. The fluctuations of these reference masses will ultimately limit the accuracy of coordinate measurements. Therefore, in a double-pumped FP cavity DFI response is not strictly speaking displacement-noise-free: we are able to cancel noise of the mirrors (the major ``headache'' in conventional detectors) but not the one of auxiliary optics (lasers, detectors, beamsplitters, etc). Another drawback of the double-pumped FP DFI is the inability to cancel laser noise since the single-cavity-scheme is not balanced. Although one may propose several FP-based balanced optical setups, for instance, modifications of conventional Michelson/Fabry-Perot topology, in all such schemes DFI response will be still limited by residual displacement noise of some test mass(es). This is a fundamental restriction of the model in Ref. \cite{2008_DFI_FP_toy_model}.

In spite of the undoubtful theoretical ``beauty'' of DFI idea proposed by S. Kawamura \textit{et al.}, such a detector, if realized in a full-scale experiment, will surely face certain technical difficulties as pointed out in Refs. \cite{2006_DTNF_GW_detection,2006_interferometers_DNF_GW_detection}. For instance, it is hardly possible to have ideal 50/50 beamsplitters or lossless mirrors. Ultimately, one may expect $100\div 1000$-fold reduction of displacement noise. In other words, although completely displacement-noise-free in theory, DFIs of the 1st type will suffer from various technical limitations in practice that will eventually leave a certain level of residual displacement noise. From this viewpoint the theoretical (fundamental) limitations of the 2nd type (incomplete) DFI schemes may not seem so dramatic as one could expect. Eventually, in both cases an experimentalist would have to deal with certain technical challenges and the only thing that matters is the total level of noise left, regardless its nature. And though suppression of residual displacement noise in DFIs of the 2nd type may turn a more complex problem than that in DFIs of the 1st type, the former still have a significant advantage --- strong GW response at low frequencies.

Therefore, it is reasonable to consider such GW detectors that would have as simple topologies as possible simultaneously allowing as complete displacement noise cancelation as possible and having the strongest response to GWs that general relativity allows. The latter requirement can be derived straightforwardly. According to relativity principle absolute coordinate and velocity measurements are prohibited. This is another formulation of indistinguishability between the GW and non-GW forces at the 0th and 1st orders of $L/\lambda_{\textrm{gw}}$. The difference rises only in the $O[h(L/\lambda_{\textrm{gw}})^2]$ order at low frequencies, as described above. If $L=c\tau$ and $\lambda_{\textrm{gw}}=c/f_{\textrm{gw}}$ then such a displacement-noise-free GW response will be proportional to $(f_{\textrm{gw}}\tau)^2h$ in spectral domain. In addition, this response can be further amplified with a FP cavity, for instance. Then one should expect the rise of the $(\gamma\tau)^{-1}\gg1$ resonant multiplier, where $\gamma$ is the cavity half-bandwidth. Ultimately, the strongest DFI GW response allowed by the first principles should be proportional to $(f_{\textrm{gw}}/\gamma)(f_{\textrm{gw}}\tau)h$ in spectral domain.

The optical setup satisfying the requirements of practical reasonableness and maximum completeness of displacement noise elimination, also restricted by condition of the strongest response derived above, does not, however, immediately follows from some basic principles. It is a matter of search at large, limited by practically reasonable configurations and assumptions. For instance, in this paper introducing several model assumptions we propose a pair of symmetrically positioned Michelson interferometers with Fabry-Perot cavities inserted into each arm of both interferometers, as a DFI GW detector with a reasonably simple optical setup and the strongest possible response.

First, consider a conventional LIGO topology (without power- and signal-recycling mirrors). Let the end mirrors be partially transmittible. In this case an interferometer will produce three response signals: the reflected (laser-noise-free) one in conventional dark park and the transmitted ones in the arms. It is worth noting here that certain care is required when calculating the responses: each response signal should be evaluated in the proper reference frame of the detector that detects the corresponding signal \cite{2008_accel_observ}. Otherwise, unmeasurable quantities may arise in the analysis. An experimentalist is able to measure the quadrature components of the interferometer responses and record them for further processing. From the set of transmitted signals quadratures it is possible, in principle, to construct a laser-noise-free linear combination. Therefore, at this stage we may obtain two signals (quadratures) free from laser phase noise: the reflected one and the combined transmitted one.

Due to the sophisticated frequency dependence of the FP cavities responses these two signals can be combined in turn to eliminate one of four differential mechanical degrees of freedoms associated with the test masses (beamsplitter, two input mirrors, two end mirrors and two end detectors). At this stage we introduce some restrictions into the optical scheme: (i) end detectors and end mirrors, and (ii) input mirrors and beamsplitter are assumed to be rigidly connected. The practical legitimacy of these assumptions seems questionable and is open for criticism, although, no basic principles forbid such a gedanken (thought) experiment. Under these restrictions we are left with only two differential degrees of freedom, one of which can be eliminated in a combination of two laser-noise-free signals. We choose to cancel displacement noise associated with the differential motion of the end platforms.

Ultimately, due to the symmetry of plane GW wavefront we are able to cancel the fluctuations of the central platform (with beamsplitter and input mirrors) if the similar interferometer is positioned symmetrically (see Fig. \ref{pic_double_Michelson_FP} in the text) and both interferometers have common central platform (this is another gedanken-experiment-supposion). In this case the single-interferometer partial DFI responses will have GW term of the same sign but the fluctuations of the central platform will enter with different signs. Adding two single-interferometer responses we cancel displacement noise of the latter. Then the obtained laser- and displacement-noise-free GW response signal turns out to be proportional to $(f_{\textrm{gw}}\tau)^2$ amplified with the cavity resonant gain $(\gamma\tau)^{-1}$.

\section{Space-time of accelerated observer in the gravitational-wave field}
\label{sec_space-time} Let us first remind briefly the ``tools'' necessary for our further considerations.

As explained in Ref. \cite{2008_accel_observ}, to obtain physically reasonable, i.e. measurable, quantities, calculations should be performed in the proper reference frames of the devices that produce the corresponding experimentally observed quantities. Since they are usually subjected to external fluctuative forces, they commit random motions and thus we have to deal with their proper non-inertial reference frames. Corresponding tools for solving certain boundary electrodynamical problems in such reference frames have been developed in Ref. \cite{2008_accel_observ} and utilized in Ref. \cite{2008_DFI_FP_toy_model}. Therefore, in this paper we will not retell the content of these works in detail but will write several useful formulas in this section.

In particular, the space-time of an observer having non-geodesic acceleration $\ddot{\xi}_{\textrm{obs}}(t)$ along the $x$-axis and falling in the weak, plane, '+'-polarized GW $h=h(t-z/c)$ propagating along the $z$-axis takes the following form:
\begin{align}
    ds^2=&-(c\,dt)^2\left[1+\frac{2}{c^2}\,\ddot{\xi}_{\textrm{obs}}(t)x\right]+
    dx^2+dy^2+dz^2\nonumber\\
    &+\,\frac{1}{2}\,\frac{x^2-y^2}{c^2}\,\ddot{h}(t-z/c)\,
    (c\,dt-dz)^2.
    \label{eq_metric_tensor}
\end{align}
Conditions of linearized theory $|2\ddot{\xi}_{\textrm{obs}}x/c^2|\ll1$ and $|h|\ll 1$ are assumed to be satisfied for all reasonable $x$ and $t$. Without the loss of generality we may assume $y=z=0$ when considering one-dimensional problems.

Consider a test mass which in a state of rest (no fluctuations and no GW) has the coordinate $x_0$ with respect to the observer (also in a state of rest). If the test mass is subjected to some external fluactuative force which moves it according to the motion law $\xi_{\textrm{t.m.}}(t)$ as seen from the laboratory frame (for instance, the one associated with the Earth surface), then its motion law with respect to the observer in space-time (\ref{eq_metric_tensor}) is:
\begin{subequations}
\begin{align}
    x_{\textrm{t.m.}}(t)&=x_0+\delta x_{\textrm{t.m.}}(t),\label{eq_law_of_motion_1}\\
    \delta x_{\textrm{t.m.}}(t)&=\frac{1}{2}\,x_0h(t)+\xi_{\textrm{t.m.}}(t)-
    \xi_{\textrm{obs}}(t),\label{eq_law_of_motion_2}
\end{align}
\end{subequations}
where $\xi_{\textrm{obs}}(t)$ is the observer's law of motion as seen from the same laboratory frame. It is assumed here that $|\xi_{\textrm{t.m.}}|,\ |\xi_{\textrm{obs}}|\ll|x_0|$. In fact, Eq. (\ref{eq_law_of_motion_2}) is the coordinate transformation from laboratory frame to the observer's frame. In spectral domain which will be widely used
\begin{equation*}
    \begin{bmatrix}
        \delta x_i(t)\\
        \xi_i(t)\\
    \end{bmatrix}=
        \int_{-\infty}^{+\infty}
    \begin{bmatrix}
        \delta x_i(\Omega)\\
        \xi_i(\Omega)\\
    \end{bmatrix}e^{-i\Omega t}\,\frac{d\Omega}{2\pi}.
\end{equation*}

It is also important to take into account the effects imposed by the GW and acceleration fields on the electromagnetic waves propagating in space-time (\ref{eq_metric_tensor}). It has been derived in Refs. \cite{2007_GW_FP_LL,2008_accel_observ} that the waves propagating in the positive and negative directions of the $x$-axis can be described by the following vector potentials:
\begin{align}
    A_\pm(x,t)=
    &A_{\pm0}\Bigl[1+g_{\pm}(x,t)+w_{\pm}(x,t)\Bigr]e^{-i(\omega_0t\mp k_0x)}\nonumber\\
    &+a_\pm(x,t)e^{-i(\omega_0t\mp k_0x)},\label{eq_EMW}
\end{align}
where
\begin{equation*}
    a_\pm(x,t)=\int_{-\infty}^{+\infty}a_\pm(\Omega+\omega_0)e^{-i\Omega
    t}\,\frac{d\Omega}{2\pi},
\end{equation*}
\begin{align*}
    g_\pm(x,t)=ik_0\Bigl[\frac{1}{4}\,x\dot{h}(t)\,\frac{x}{c}\mp
    \frac{1}{2}\,xh(t)
    +\frac{c}{2}\int_{t\mp\frac{x}{c}}^{t}h(t_1)dt_1\Bigr],
\end{align*}
and
\begin{equation*}
    w_\pm(x,t)=ik_0\Bigl[-\dot{\xi}_{\textrm{obs}}(t)\,\frac{x}{c}\pm
    \xi_{\textrm{obs}}(t)\mp\xi_{\textrm{obs}}(t\mp x/c)\Bigr].
\end{equation*}
Both $g_\pm(x,t)$ and $w_\pm(x,t)$ describe the distributed effects: $g_\pm$ is responsible for the direct coupling between the GW and the electromagnetic wave and $w_\pm$ describe the redshift imposed on the electromagnetic wave by the noninertiality of the reference frame. Weak fields $a_\pm(x,t)$ describe electromagnetic fluctuations (classical or quantum).

In this paper we will also deal with the motions along the $y$-axis. In this case all the formulas remain the same but the GW function $h(t)$ should be taken with the opposite sign (this follows from the metric (\ref{eq_metric_tensor})).

\section{Responses of a Fabry-Perot cavity to the gravitational wave}\label{sec_FP_responses} 
\begin{figure}[h]
\begin{center}
\includegraphics[scale=0.65]{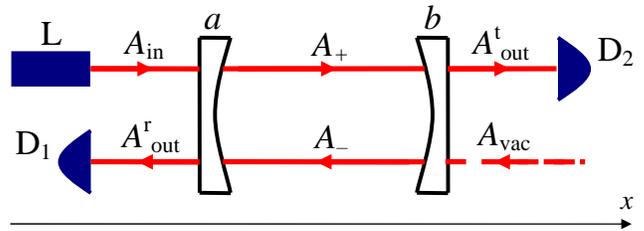}
\caption{Fabry-Perot cavity assembled of two movable mirrors $a$ and
$b$. Cavity is pumped by laser L through mirror $a$ with the input
wave $A_{\textrm{in}}(x,t)$ and through mirror $b$ with the
vacuum-state wave $A_{\textrm{vac}}(x,t)$. Optical field inside the
cavity is represented as a sum of the wave $A_{+}(x,t)$, running in
the positive direction of the $x$-axis, and the wave $A_{-}(x,t)$,
running in the opposite direction. The wave reflected from the
cavity $A_{\textrm{out}}^{\textrm{r}}(x,t)$ is measured with the
(amplitude or balanced homodyne) detector $\textrm{D}_1$ with the
reference oscillation (if necessary) produced by laser L.
Transmitted wave $A_{\textrm{out}}^{\textrm{t}}(x,t)$ is measured
with the (amplitude or balanced homodyne) detector $\textrm{D}_2$
with the reference oscillation (if necessary) produced by some local
source.} \label{pic_FP_cavity}
\end{center}
\end{figure}

Let us consider the operation of a single Fabry-Perot cavity as a GW detector (see Fig. \ref{pic_FP_cavity}).
Cavity is assembled of two movable mirrors $a$ and $b$, both lossless and having the amplitude transmission coefficient $T$, $|T|\ll1$. We put distance between the mirrors in the absence of the gravitational wave and optical radiation to be equal to $L$. The incident GW $h(t-z/c)$ is assumed to be weak, plane, '+'-polarized and propagating along the $z$-axis. Then without the loss of generality we assume the cavity to be lying in the plane $z=0$ along one of the GW principal axes, coinciding with the $x$-axis.

Cavity is pumped by laser L whose center of mass commits a fluctuative motion $\xi_{\textrm{L}}(t)$ along the $x$-axis as seen from the laboratory frame. Both mirrors $a$ and $b$ have associated displacement noise $\xi_a(t)$ and $\xi_b(t)$. Finally, both the detectors $\textrm{D}_1$ and $\textrm{D}_2$ that measure the reflected wave and transmitted wave correspondingly fluctuate as $\xi_{\textrm{D}_1}(t)$ and $\xi_{\textrm{D}_2}(t)$.

To evaluate the response signals of the cavity we should perform the calculations in the proper reference frames of detectors $\textrm{D}_1$ and $\textrm{D}_2$, as pointed above: the reflected signal is measured with the first one and the transmitted signal is measured with the latter one.

\subsection{Evaluation of the reflected signal}
Here we will derive the expression for the wave reflected from the cavity. Since it is detected by $\textrm{D}_1$ we will perform the calculation its proper reference frame. This section mostly repeats the similar considerations in Ref. \cite{2008_DFI_FP_toy_model}, therefore we will proceed without detailed comments.

The origin of the coordinate system is assumed to be set up at the center of mass of $\textrm{D}_1$. Therefore, according to Eqs. (\ref{eq_law_of_motion_1},\ref{eq_law_of_motion_2}) test masses of the system will have the following motion laws with respect to $\textrm{D}_1$:
\begin{align*}
    x_{\textrm{D}_1}(t)&=\xi_{\textrm{D}_1}(t)-\xi_{\textrm{D}_1}(t)=0,\\
    x_\textrm{L}(t)&=\xi_\textrm{L}(t)-\xi_{\textrm{D}_1}(t),\\
    x_a(t)&\approx\xi_a(t)-\xi_{\textrm{D}_1}(t),\\
    x_b(t)&\approx L+\frac{1}{2}\,Lh(t)+\xi_b(t)-\xi_{\textrm{D}_1}(t).
\end{align*}
In last two equations we neglected the small distance (compared to the cavity length) between the optical bench where laser and detector $\textrm{D}_1$ are located and the input mirror.

Let the cavity be pumped by laser L through the input mirror $a$ with the input wave
\begin{align}
    A_{\textrm{in}}(x,t)&=
    A_{\textrm{in}0}\Bigl[1+g_+(x,t)+w_+(x,t)\Bigr]\nonumber\\
    &\quad\times\exp\Biggl\{-i\omega_0\left[t-
    \frac{x-x_\textrm{L}(t)}{c}\right]\Biggr\}\nonumber\\
    &\quad+a_{\textrm{in}}(x,t)e^{-i(\omega_0t-k_0x)},\label{eq_input_wave_R}
\end{align}
Strictly speaking, the argument of $x_\textrm{L}$ here should depend on $x_\textrm{L}$ itself like $t-x_\textrm{L}/c$, but since $x_\textrm{L}$ is already the quantity of the 1st order of smallness we can neglect such dependence. The vacuum-state pump through mirror $b$ can be written as:
\begin{equation}
    A_{\textrm{vac}}(x,t)=a_{\textrm{vac}}(x,t)e^{-i\bigl[\omega_0t+k_0(x-L)\bigr]}.
    \label{eq_vacuum_wave_R}
\end{equation}
Here $a_{\textrm{in}}(x,t)$ is the ``weak'' field describing optical laser noise of the pump wave and $a_{\textrm{vac}}(x,t)$ is the ``weak'' field describing vacuum noise in the opposite input port. Remind, that both the laser and mirror $a$ are located at $x\approx0$, where $g_+(0,t)=w_+(0,t)=0$, thus input wave does not acquire distributed phase shift when it reaches mirror $a$.

It is convenient to represent the optical field inside the cavity as a sum of two waves, $A_{+}(x,t)$ and $A_{-}(x,t)$, running in the opposite directions:
\begin{align}
    A_{\pm}(x,t)=&A_{\pm0}\Bigl[1+g_\pm(x,t)+w_\pm(x,t)\Bigr]
    e^{-i(\omega_0t\mp k_0x)}\nonumber\\
    &+a_{\pm}(x,t)e^{-i(\omega_0t\mp k_0x)}.\label{eq_inside_wave_R}
\end{align}
Here $a_\pm(x,t)$ describe the phase shift accumulated by the light wave while circulating inside the cavity.

Output wave reflected from the cavity is:
\begin{align}
    A^{\textrm{r}}_{\textrm{out}}(x,t)
    =&A^{\textrm{r}}_{\textrm{out}0}\Bigl[1+g_-(x,t)+w_-(x,t)\Bigr]
    e^{-i(\omega_0t+k_0x)}\nonumber\\
    &+a^{\textrm{r}}_{\textrm{out}}(x,t)e^{-i(\omega_0t+k_0x)},
    \label{eq_reflected_wave}
\end{align}
If detector $\textrm{D}_1$ is a quadratic amplitude detector then it measures the quantity proportional to $|A^{\textrm{r}}_{\textrm{out}}(x,t)|^2$ (neglecting very small terms of the order of $\Omega/\omega_0$). If detector is a balanced homodyne detector then it measures the quadratures of $A^{\textrm{r}}_{\textrm{out}}(x,t)$. In this case the reference oscillation can be produced by laser L. We will call $a^{\textrm{r}}_{\textrm{out}}(x,t)$ the reflected signal below.

To obtain the reflected signal we substitute fields (\ref{eq_input_wave_R} -- \ref{eq_reflected_wave}) into the set of boundary conditions (conditions of the electric field continuity along the surfaces of the mirrors) \cite{2007_GW_FP_LL,2008_DFI_FP_toy_model}:
\begin{align*}
    A_+(x_a(t),t)&=TA_{\textrm{in}}(x_a(t),t)-RA_-(x_a(t),t),\\
    A_-(x_b(t),t)&=TA_{\textrm{vac}}(x_b(t),t)-RA_+(x_b(t),t),\\
    A^{\textrm{r}}_{\textrm{out}}(x_a(t),t)&=RA_{\textrm{in}}(x_a(t),t)+TA_-(x_a(t),t),
\end{align*}
and solve the system with the method of successive approximations (see Ref. \cite{2008_DFI_FP_toy_model}). The required solution of the 1st order in spectral domain is:
\begin{align}
    a^{\textrm{r}}_{\textrm{out}}&=
    \mathcal{R}(a_{\textrm{in}}-A_{\textrm{in}0}ik_0\xi_{\textrm{L}})+
    \mathcal{T}a_{\textrm{vac}}\nonumber\\
    &\quad+TA_{-0}2ik_0\,
    \frac{(\xi_b+\xi^{\textrm{r.t.}}_{\textrm{gw}})e^{i\Omega\tau}-\xi_a}
    {\mathcal{T}^2_{\omega_0+\Omega}}\nonumber\\
    &\quad+A^{\textrm{r}}_{\textrm{out}0}2ik_0\xi_a-
    A^{\textrm{r}}_{\textrm{out}0}ik_0\xi_{\textrm{D}_1}.
    \label{eq_reflected_signal_FP}
\end{align}
The following notations have been introduced:
\begin{align*}
    \mathcal{T}^2_{\omega_0}&=1-R^2e^{2i\omega_0\tau},\qquad
    \mathcal{T}^2_{\omega_0+\Omega}=1-R^2e^{2i(\omega_0+\Omega)\tau},\\
    \mathcal{R}&=
    \frac{R-Re^{2i(\omega_0+\Omega)\tau}}{1-R^2e^{2i(\omega_0+\Omega)\tau}},\qquad
    \mathcal{T}=\frac{T^2e^{i(\omega_0+\Omega)\tau}}{1-R^2e^{2i(\omega_0+\Omega)\tau}}\,
    A_{\textrm{in}0},\\
    A_{-0}&=-\frac{RTe^{2i\omega_0\tau}}{\mathcal{T}^2_{\omega_0}}\,A_{\textrm{in}0},\qquad
    A^{\textrm{r}}_{\textrm{out}0}
    =\frac{R-Re^{2i\omega_0\tau}}{\mathcal{T}^2_{\omega_0}}\,A_{\textrm{in}0},\\
    \xi^{\textrm{r.t.}}_{\textrm{gw}}&=\frac{1}{2}\,
    Lh\,\frac{\sin\Omega\tau}{\Omega\tau},
\end{align*}
having the following physical meaning: $1/\mathcal{T}^2_{\omega_0}$ describes the resonant amplification of the input amplitude $A_{\textrm{in}0}$ inside the cavity, $1/\mathcal{T}^2_{\omega_0+\Omega}$ describes the frequency-dependent resonant amplification of the variation of the circulating light wave, $\mathcal{R}$ and $\mathcal{T}$ are the generalized coefficients of reflection (from a FP cavity) and transmission (through a FP cavity), $A_{-0}$ is the mean amplitude of the optical wave inside the cavity running in the negative direction of the $x$-axis, $A^{\textrm{r}}_{\textrm{out}0}$ is the mean amplitude of the wave reflected from the cavity and $\xi^{\textrm{r.t.}}_{\textrm{gw}}$ is the response to GW after a single round trip of light inside the cavity

It is also very useful to analyze the physical meaning of each summand in formula (\ref{eq_reflected_signal_FP}):
\begin{enumerate}
    \item{$\mathcal{R}(a_{\textrm{in}}-A_{\textrm{in}0}ik_0\xi_{\textrm{L}})$. This term states that the optical laser noise $a_{\textrm{in}}$ is indistinguishable from laser displacement noise $\xi_{\textrm{L}}$ so both always come together and their sum is usually called laser phase noise. In spectral domain reflected wave obviously contains laser phase noise multiplied by the generalized coefficient of reflection $\mathcal{R}$.}
    \item{$\mathcal{T}a_{\textrm{vac}}$ is the vacuum noise $a_{\textrm{vac}}$ from the opposite input port which is transmitted through the cavity and comes with the corresponding coefficient of transmission $\mathcal{T}$ in the reflected signal.}
    \item{$TA_{-0}2ik_0\Bigl[(\xi_b+\xi_{\textrm{gw}})e^{i\Omega\tau}-\xi_a\Bigr]/\mathcal{T}^2_{\omega_0+\Omega}$ describes the part of reflected wave which flows out of the cavity and, obviously, represents the phase shift accumulated by optical wave circulating inside the cavity. This phase shift is proportional to the change of cavity length due to fluctuative motion of the mirrors $\xi_{a,b}$ and the change of optical path due to GW $\xi_{\textrm{gw}}$. Accumulation of the signal is described by the resonant factor $\mathcal{T}^2_{\omega_0+\Omega}$.}
    \item{$A^{\textrm{r}}_{\textrm{out}0}2ik_0\xi_a$ is the phase shift due to the prompt reflection from the input mirror.}
    \item{$A^{\textrm{r}}_{\textrm{out}0}ik_0\xi_{\textrm{D}_1}$ accounts for the motion of detector which receives the reflected signal. It is straightforward to verify that this summand will be canceled in the amplitude detection, evidently meaning that the amplitude detector is insusceptible to its own motion (since amplitude is measured, but not the phase).}
\end{enumerate}
It is worth noting that the same formula (\ref{eq_reflected_signal_FP}) one can derive without resorting to the non-inertial frame by linearly combining the considerations in (i) the transverse-traceless gauge for the GW response and (ii) the laboratory frame for response to the fluctuations of test masses. Such a simplification owns to the specific feature of a round-trip scheme: the source and the receiver of optical radiation are located approximately at the same spacial position and thus their clocks could be synchronized almost perfectly.

\subsection{Evaluation of transmitted signal}
Here we will derive the expression for the transmitted signal. Since the source L and the receiver $\textrm{D}_2$ are separated by a large distance, the effect of redshift due to the acceleration of detector arises which cannot be obtained solely in the laboratory frame. Therefore, we are forced to perform the calculations in the proper reference frame of detector $\textrm{D}_2$.

We place the origin of the coordinates at the center of mass of $\textrm{D}_2$. Then according to Eqs. (\ref{eq_law_of_motion_1}) and (\ref{eq_law_of_motion_2}) other test masses move as following:
\begin{align*}
    x_{\textrm{D}_2}(t)&=\xi_{\textrm{D}_2}(t)-\xi_{\textrm{D}_2}(t)=0,\\
    x_\textrm{L}(t)&\approx-L-\frac{1}{2}\,Lh(t)+\xi_\textrm{L}(t)-\xi_{\textrm{D}_2}(t),\\
    x_a(t)&\approx -L-\frac{1}{2}\,Lh(t)+\xi_a(t)-\xi_{\textrm{D}_2}(t),\\
    x_b(t)&\approx \xi_b(t)-\xi_{\textrm{D}_2}(t).
\end{align*}
Again we neglected the small constant distances separating laser L and input mirror $a$, and mirror $b$ and detector $\textrm{D}_2$.

From the viewpoint of detector $\textrm{D}_2$ laser L emits the input wave described by the following vector potential:
\begin{align}
    A_{\textrm{in}}(x,t)&=A_{\textrm{in}0}
    \Biggl[1+g_+(x,t)-g_+\left(-L,t-\frac{x+L}{c}\right)\nonumber\\
    &\qquad\qquad+w_+(x,t)-w_+\left(-L,t-\frac{x+L}{c}\right)\Biggr]\nonumber\\
    &\qquad\quad\times\exp\left\{-i\omega_0\left[t-\frac{x-
    x_\textrm{L}\left(t-\frac{x+L}{c}\right)}{c}\right]\right\}\nonumber\\
    &\quad+a_{\textrm{in}}(x,t)e^{-i\bigl[\omega_0t-k_0(x+L)\bigr]}
    \label{eq_input_wave_T}
\end{align}
This corresponds to the condition that in the immediate vicinity of the source optical wave acquires neither phase shift due to the motion of the source, nor phase shift due to both distributed effects.

The vacuum pump is:
\begin{equation}
    A_{\textrm{vac}}(x,t)=a_{\textrm{vac}}(x,t)e^{-i(\omega_0t+k_0x)}.
    \label{eq_vacuum_wave_T}
\end{equation}

It is most convenient to write the wave traveling in the positive direction of the $x$-axis inside the cavity as following:
\begin{align}
    A_{+}(x,t)&=A_{+0}
    \Biggl[1+g_+(x,t)-g_+\left(-L,t-\frac{x+L}{c}\right)\nonumber\\
    &\qquad\qquad+w_+(x,t)-w_+\left(-L,t-\frac{x+L}{c}\right)\Biggr]\nonumber\\
    &\qquad\quad\times\exp\Biggl\{-i\omega_0\left[t-\frac{x+L}{c}\right]\Biggr\}\nonumber\\
    &\quad+a_{+}(x,t)e^{-i\bigl[\omega_0t-k_0(x+L)\bigr]}.
    \label{eq_positive_wave_T}
\end{align}
Such a notation means that the phase is counted from $x=-L$. The motion of the input mirror $a$ will be taken into account in the boundary conditions.

The phase of the optical wave traveling in the negative direction will be counted from $x=0$ (remind, that both $g_-(x,t)$ and $w_-(x,t)$ vanish at $x=0$):
\begin{align}
    A_-(x,t)&=A_{-0}\Bigl[1+g_-(x,t)+w_-(x,t)\Bigr]
    e^{-i(\omega_0t+k_0x)}\nonumber\\
    &\quad+a_-(x,t)e^{-i(\omega_0t+k_0x)}.\label{eq_negative_wace_T}
\end{align}

Finally, the transmitted wave with the phase counted from $x=0$ ($g_+(0,t)=w_+(0,t)=0$) is:
\begin{align}
    A_{\textrm{out}}^\textrm{t}(x,t)&=A_{\textrm{out}0}^\textrm{t}
    \Bigl[1+g_+(x,t)+w_+(x,t)\Bigr]e^{-i(\omega_0t-k_0x)}\nonumber\\
    &\quad+a_{\textrm{out}}^\textrm{t}(x,t)e^{-i(\omega_0t-k_0x)}.
    \label{eq_transmitted_wave}
\end{align}
If detector $\textrm{D}_1$ is a quadratic amplitude detector then it measures the quantity proportional to $|A^{\textrm{t}}_{\textrm{out}}(x,t)|^2$ (neglecting very small terms of the order of $\Omega/\omega_0$). If detector is a balanced homodyne detector then it measures the quadratures of $A^{\textrm{t}}_{\textrm{out}}(x,t)$. In this case the reference oscillation should be produced by some local source. We call $a^{\textrm{t}}_{\textrm{out}}(x,t)$ the transmitted signal below.

To obtain the transmitted signal we substitute fields (\ref{eq_input_wave_T} -- \ref{eq_transmitted_wave}) into the set of boundary conditions:
\begin{align*}
    A_+(x_a(t),t)&=TA_{\textrm{in}}(x_a(t),t)-RA_-(x_a(t),t),\\
    A_-(x_b(t),t)&=TA_{\textrm{vac}}(x_b(t),t)-RA_+(x_b(t),t)\\
    A_{\textrm{out}}^\textrm{t}(x_b(t),t)&=RA_{\textrm{vac}}(x_b(t),t)+TA_+(x_b(t),t).
\end{align*}
The required 1st order solution of this system in spectral domain is:
\begin{align}
    a^{\textrm{t}}_{\textrm{out}}&=
    \mathcal{T}\Bigl[a_{\textrm{in}}-
    A_{\textrm{in}0}ik_0\left(\xi_{\textrm{L}}+\xi^{\textrm{f.t.}}_{\textrm{gw}}
    -i\Omega\tau\xi_{\textrm{D}_2}\right)\Bigr]+\mathcal{R}a_{\textrm{vac}}\nonumber\\
    &\quad+R^2Te^{3i\omega_0\tau}A_{+0}2ik_0\,
    \frac{(\xi_b+\xi^{\textrm{r.t.}}_{\textrm{gw}})e^{i\Omega\tau}-\xi_a}
    {\mathcal{T}^2_{\omega_0+\Omega}}\,e^{i\Omega\tau}\nonumber\\
    &\quad+A^{\textrm{t}}_{\textrm{out}0}ik_0\xi_{\textrm{D}_2},
    \label{eq_transmitted_signal_FP}
\end{align}
where
\begin{align*}
    \xi^{\textrm{f.t.}}_{\textrm{gw}}&=\frac{1}{4}\,(-i\Omega\tau)Lh-
    \frac{1}{2}\,\frac{Lh}{i\Omega\tau}(1-e^{-i\Omega\tau}),\\
    A_{+0}&=\frac{T}{\mathcal{T}^2_{\omega_0}}\,A_{\textrm{in}0},\qquad
    A_{\textrm{out}0}^\textrm{t}=\frac{T^2e^{i\omega_0\tau}}{\mathcal{T}^2_{\omega_0}}\,
    A_{\textrm{in}0}.
\end{align*}
Here $\xi^{\textrm{f.t.}}_{\textrm{gw}}$ is the forward-trip GW response, $A_{+0}$ is the mean amplitude of the wave traveling in the positive direction of the $x$-axis inside the cavity and $A_{\textrm{out}0}^\textrm{t}$ is the mean amplitude of the transmitted wave.

Again it is useful to analyze the physical meaning of each summand in formula (\ref{eq_transmitted_signal_FP}):
\begin{enumerate}
    \item{$\mathcal{T}\Bigl[a_{\textrm{in}}-A_{\textrm{in}0}ik_0\left(\xi_{\textrm{L}}+\xi^{\textrm{f.t.}}_{\textrm{gw}}
    -i\Omega\tau\xi_{\textrm{D}_2}\right)\Bigr]$ describes laser noise transmitted through the cavity. The summand with $\xi^{\textrm{f.t.}}_{\textrm{gw}}-i\Omega\tau\xi_{\textrm{D}_2}$ means that the initial phase shift due to laser displacement is additionally redshifted with the GW and the motion of observer (detector $\textrm{D}_2$), because GW and acceleration fields change the rate of laser clock with respect to detector $\textrm{D}_2$ clock. It should be also noted that even the amplitude detector will be susceptible to $-i\Omega\tau\xi_{\textrm{D}_2}$ ($\dot{\xi}_{\textrm{D}_2}(t)\tau$ in time domain), since phase modulation is transformed into amplitude modulation in a FP cavity.}
    \item{$\mathcal{R}a_{\textrm{vac}}$ is the vacuum noise reflected from the cavity into the transmitted port.}
    \item{$R^2Te^{3i\omega_0\tau}A_{+0}2ik_0\,\Bigl[(\xi_b+\xi^{\textrm{r.t.}}_{\textrm{gw}})e^{i\Omega\tau}-\xi_a\Bigr]e^{i\Omega\tau}/
    {\mathcal{T}^2_{\omega_0+\Omega}}$ describes the total variation of the phase accumulated inside the cavity.}
    \item{$A^{\textrm{t}}_{\textrm{out}0}ik_0\xi_{\textrm{D}_2}$ accounts for the displacement of the receiver. If detector is the amplitude one, this term becomes unmeasurable.}
\end{enumerate}

\section{Michelson/Fabry-Perot topology}\label{sec_Michelson_FP}
The major disadvantage of a single cavity-based GW detector is the significant level of laser noise which dominates over other noises in practice. To cancel laser noise one should implement a balanced optical setup, for instance, a Michelson interferometer tuned to dark-port regime.

\subsection{Response signals of a Michelson/Fabry-Perot interferometer}\label{subsec_responses_Michelson_FP} 
\begin{figure}[h]
\begin{center}
\includegraphics[scale=0.57]{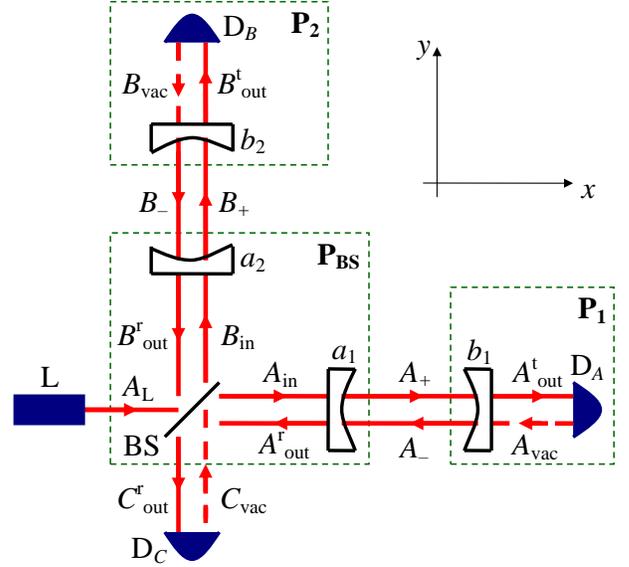}
\caption{A Michelson/Fabry-Perot optical setup. Two Fabry-Perot
cavities $a_1b_1$ and $a_2b_2$ are inserted into Michelson
interferometer horizontal and vertical arms correspondingly.
Interferometer is pumped by laser L through beamsplitter BS with the
input wave $A_{\textrm{L}}(x,t)$. Beamsplitter produces to waves
$A_{\textrm{in}}(x,t)$ and $B_{\textrm{in}}(y,t)$ which pump
horizontal and vertical cavities respectively. Interferometer is
tuned to the dark-port regime so that both reflected waves
$A_{\textrm{out}}^{\textrm{r}}(x,t)$ and
$B_{\textrm{out}}^{\textrm{r}}(y,t)$ destructively interfere at the
beamsplitter and all mean power returns towards laser L (not shown
in the Fig.). The signal part $C_{\textrm{out}}^{\textrm{r}}(y,t)$
containing the accumulated phase shift penetrates into the dark port
and is incident on detector $\textrm{D}_C$ which operates as a
balanced homodyne detector with the reference oscillation produced
by laser L. The dark port of detector $\textrm{D}_C$ also produces
the vacuum pump $C_{\textrm{vac}}(y,t)$. Transmitted waves
$A_{\textrm{out}}^{\textrm{t}}(x,t)$ and
$B_{\textrm{out}}^{\textrm{t}}(y,t)$ are detected with detectors
$\textrm{D}_A$ and $\textrm{D}_B$ correspondingly which may operate
as amplitude or balanced homodyne detectors. In the latter case
reference oscillations are produced by some local sources. Both
detector ports also produce vacuum pumps $A_{\textrm{vac}}(x,t)$ and
$B_{\textrm{vac}}(y,t)$.} \label{pic_Michelson_FP}
\end{center}
\end{figure}

Let us consider a Michelson-type interferometer with FP cavities in its arms (see Fig. \ref{pic_Michelson_FP}).
Laser L which randomly moves along the $x$-axis as $\xi_{\textrm{L}}(t)$ pumps the interferometer with the input wave $A_{\textrm{L}}(x,t)$. Upon arrival to 50/50-beamsplitter BS which may fluctuate along the $x$- and $y$-axes as $\xi_{\textrm{BS}}(t)$ and $\eta_{\textrm{BS}}(t)$ respectively, the input wave is splitted into two waves $A_{\textrm{in}}(x,t)$ and $B_{\textrm{in}}(y,t)$ which pump horizontal and vertical arms respectively. Fabry-Perot cavity in the horizontal arm is assembled of two mirrors $a_1$ and $b_1$ which fluctuate along the $x$-axis as $\xi_{a_1}(t)$ and $\xi_{b_1}(t)$. The similar cavity in the vertical arm is assembled of mirrors $a_2$ and $b_2$ which fluctuate along the $y$-axis as $\eta_{a_2}(t)$ and $\eta_{b_2}(t)$. Both the cavities may produce reflected and transmitted waves. Reflected waves $A_{\textrm{out}}^{\textrm{r}}(x,t)$ and $B_{\textrm{out}}^{\textrm{r}}(y,t)$ return towards the beamsplitter and interfere. Assume the interferometer is tuned to dark port regime. This means that the reflected waves interfere destructively and the mean optical power returns towards the laser. However, the weak time-dependent (signal) part $C_{\textrm{out}}^{\textrm{r}}(y,t)$ penetrates into the dark port and falls on detector $\textrm{D}_C$ which fluctuates along the $y$-axis as $\eta_{\textrm{D}_C}(t)$. Let the latter one operate as balanced homodyne detector with the reference oscillation produced by laser L. Dark port of $\textrm{D}_C$ also produces the vacuum pump $C_{\textrm{vac}}(y,t)$. Transmitted waves $A_{\textrm{out}}^{\textrm{t}}(x,t)$ and $B_{\textrm{out}}^{\textrm{t}}(y,t)$ are measured with the corresponding detectors: in horizontal arm it is $\textrm{D}_A$ and in vertical arm it is $\textrm{D}_B$ moving randomly as $\xi_{\textrm{D}_A}(t)$ and $\eta_{\textrm{D}_B}(t)$ correspondingly. Both detectors may operate either as amplitude or homodyne ones. In the latter case reference oscillations should be produced by some local sources (see below). Both detector ports also produce vacuum pumps $A_{\textrm{vac}}(x,t)$ and $B_{\textrm{vac}}(y,t)$.

Let us now write the equations for input and output waves in an interferometer. We will not write the expressions for the waves explicitly; one can obtain them straightforwardly making obvious changes in the formulas from the previous section. At beamsplitter the relation between the input waves is as following:
\begin{align*}
    A_{\textrm{in}}(x_{\textrm{BS}}(t),t)&=
    \frac{1}{\sqrt{2}}\,A_{\textrm{L}}(x_{\textrm{BS}}(t),t)-
    \frac{1}{\sqrt{2}}\,C_{\textrm{vac}}(y_{\textrm{BS}}(t),t),\nonumber\\
    B_{\textrm{in}}(y_{\textrm{BS}}(t),t)&=
    \frac{1}{\sqrt{2}}\,A_{\textrm{L}}(x_{\textrm{BS}}(t),t)+
    \frac{1}{\sqrt{2}}\,C_{\textrm{vac}}(y_{\textrm{BS}}(t),t).
\end{align*}
The relation between the reflected waves is:
\begin{equation*}
    C_{\textrm{out}}^{\textrm{r}}(y_{\textrm{BS}}(t),t)=
    \frac{1}{\sqrt{2}}\,B_{\textrm{out}}^{\textrm{r}}(y_{\textrm{BS}}(t),t)-
    \frac{1}{\sqrt{2}}\,A_{\textrm{out}}^{\textrm{r}}(x_{\textrm{BS}}(t),t).
\end{equation*}
At this stage we do not define the reference frame, therefore, fields and coordinates of the test masses should be specified explicitly for this or that frame.

These equations can be solved straightforwardly. However, we do not need to do this since we already know the solution for a single cavity (\ref{eq_reflected_signal_FP}). First, we need to write explicitly the expressions for weak reflected fields $a^{\textrm{r}}_{\textrm{out}}$ and $b^{\textrm{r}}_{\textrm{out}}$. To do this we use the first pair of BS boundary conditions to obtain in spectral domain:
\begin{subequations}
\begin{align}
    A_{\textrm{in}0}&=\frac{1}{\sqrt{2}}\,A_{\textrm{L}0},\qquad
    B_{\textrm{in}0}=\frac{1}{\sqrt{2}}\,A_{\textrm{L}0},\label{eq_mean_amplitudes}\\
    a_{\textrm{in}}&=\frac{1}{\sqrt{2}}\,a_{\textrm{L}}-
    \frac{1}{\sqrt{2}}\,c_{\textrm{vac}},\label{eq_input_wave_horizontal}\\
    b_{\textrm{in}}&=\frac{1}{\sqrt{2}}\,a_{\textrm{L}}+
    \frac{1}{\sqrt{2}}\,c_{\textrm{vac}}+
    \frac{1}{\sqrt{2}}\,A_{\textrm{L}0}ik_0(\delta x_{\textrm{BS}}-\delta y_{\textrm{BS}}),
    \label{eq_input_wave_vertical}
\end{align}
\end{subequations}
Now let us specify the reference frame. Since both reflected waves will ultimately end up at detector $\textrm{D}_C$, it is necessary to work in its proper frame. Another way is to use the laboratory frame which implementation is justified with the round-trip situation: laser L and detector $\textrm{D}_C$ can be approximately considered as located at the same spatial position. In any case, substitution Eqs. (\ref{eq_mean_amplitudes} -- \ref{eq_input_wave_vertical}) into formula (\ref{eq_reflected_signal_FP}) we obtain:
\begin{align*}
    a^{\textrm{r}}_{\textrm{out}}&=
    \frac{1}{\sqrt{2}}\,\mathcal{R}(a_{\textrm{L}}-c_{\textrm{vac}}-
    A_{\textrm{L}0}ik_0\xi_{\textrm{L}})+\mathcal{T}a_{\textrm{vac}}\nonumber\\
    &\quad+TA_{-0}2ik_0\,
    \frac{(\xi_{b_1}+\xi^{\textrm{r.t.}}_{\textrm{gw}})e^{i\Omega\tau}-\xi_{a_1}}
    {\mathcal{T}^2_{\omega_0+\Omega}}+A^{\textrm{r}}_{\textrm{out}0}2ik_0\xi_{a_1},\\
    %%%%%%%%%%%%%%%%%%%%%%%%%%%%%%%%%%%%%%%%%%%%%%%%%%%%%%%%%%%%%%%%%%%%%%%%%%%%%%%
    b^{\textrm{r}}_{\textrm{out}}&=
    \frac{1}{\sqrt{2}}\,\mathcal{R}\bigl[a_{\textrm{L}}+c_{\textrm{vac}}-
    A_{\textrm{L}0}ik_0(\xi_{\textrm{L}}-\xi_{\textrm{BS}}+\eta_{\textrm{BS}})\bigr]+
    \mathcal{T}b_{\textrm{vac}}\nonumber\\
    &\quad+TB_{-0}2ik_0\,
    \frac{(\xi_{b_2}-\xi^{\textrm{r.t.}}_{\textrm{gw}})e^{i\Omega\tau}-\xi_{a_2}}
    {\mathcal{T}^2_{\omega_0+\Omega}}+B^{\textrm{r}}_{\textrm{out}0}2ik_0\xi_{a_2}.
\end{align*}
Here we assumed that the '+'-polarized GW is perfectly aligned along the interferometer arms. We also neglected the term proportional to $\eta_{\textrm{D}_C}$ since the signal wave does not include ``strong'' mean component. For simplicity let us assume that both the cavities have equal detunings and bandwidths. This results in $A_{-0}=B_{-0}$ and $A^{\textrm{r}}_{\textrm{out}0}=B^{\textrm{r}}_{\textrm{out}0}$.

The boundary condition for reflected waves dictates that:
\begin{align*}
    C^{\textrm{r}}_{\textrm{out}0}&=\frac{1}{\sqrt{2}}\,B^{\textrm{r}}_{\textrm{out}0}-
    \frac{1}{\sqrt{2}}\,A^{\textrm{r}}_{\textrm{out}0},\nonumber\\
    c^{\textrm{r}}_{\textrm{out}}&=\frac{1}{\sqrt{2}}\,b^{\textrm{r}}_{\textrm{out}}-
    \frac{1}{\sqrt{2}}\,a^{\textrm{r}}_{\textrm{out}}+
    \frac{1}{\sqrt{2}}\,A^{\textrm{r}}_{\textrm{out}0}ik_0
    (\delta x_{\textrm{BS}}-\delta y_{\textrm{BS}}).
\end{align*}
Substituting here the obtained expressions for $a^{\textrm{r}}_{\textrm{out}}$ and $b^{\textrm{r}}_{\textrm{out}}$ we obtain:
%\begin{align*}
%    c^{\textrm{r}}_{\textrm{out}}&=\mathcal{R}c_{\textrm{vac}}+
%    \frac{1}{\sqrt{2}}\,\mathcal{T}(b_{vac}-a_{\textrm{vac}})\nonumber\\
%    &\quad-\frac{1}{\sqrt{2}}\,TA_{-0}2ik_0\,
%    \frac{\xi_{b_1}-\eta_{b_2}+2\xi^{\textrm{r.t.}}_{\textrm{gw}}}
%    {\mathcal{T}^2_{\omega_0+\Omega}}\,e^{i\Omega\tau}\nonumber\\
%    &\quad+\frac{1}{\sqrt{2}}\,TA_{-0}2ik_0\,
%    \frac{\xi_{a_1}-\eta_{a_2}}{\mathcal{T}^2_{\omega_0+\Omega}}
%    -\frac{1}{\sqrt{2}}\,A^{\textrm{r}}_{\textrm{out}0}2ik_0(\xi_{a_1}-\eta_{a_2})\nonumber\\
%    &\quad+\frac{1}{\sqrt{2}}\,A^{\textrm{r}}_{\textrm{out}0}2ik_0
%    (\xi_{\textrm{BS}}-\eta_{\textrm{BS}})+
%    \frac{1}{2}\,\mathcal{R}A_{\textrm{L}0}ik_0(\xi_{\textrm{BS}}-\eta_{\textrm{BS}}).
%\end{align*}
\begin{align}
    c^{\textrm{r}}_{\textrm{out}}&=\mathcal{R}c_{\textrm{vac}}+
    \frac{1}{\sqrt{2}}\,\mathcal{T}(b_{\textrm{vac}}-a_{\textrm{vac}})\nonumber\\
    &\quad-\frac{1}{\sqrt{2}}\,TA_{-0}2ik_0\,
    \frac{\xi_{b_1}-\eta_{b_2}+2\xi^{\textrm{r.t.}}_{\textrm{gw}}}
    {\mathcal{T}^2_{\omega_0+\Omega}}\,e^{i\Omega\tau}\nonumber\\
    &\quad+\frac{1}{\sqrt{2}}\left[\frac{TA_{-0}}{\mathcal{T}^2_{\omega_0+\Omega}}
    -A^{\textrm{r}}_{\textrm{out}0}\right]2ik_0(\xi_{a_1}-\eta_{a_2})\nonumber\\
    &\quad+\frac{1}{\sqrt{2}}\bigl[A^{\textrm{r}}_{\textrm{out}0}+
    \mathcal{R}A_{\textrm{L}0}\bigr]ik_0(\xi_{\textrm{BS}}-\eta_{\textrm{BS}}).
    \label{eq_reflected_signal_Michelson}
\end{align}
One can note that the obtained signal is very similar to the one of a single cavity. Namely, the following degrees of freedom are equivalent from this viewpoint: $\xi_a\leftrightarrow\eta_{a_2}-\xi_{a_1}$, $\xi_b\leftrightarrow\eta_{b_2}-\xi_{b_1}$, $\xi_\textrm{L}\leftrightarrow\eta_{\textrm{BS}}-\xi_{\textrm{BS}}$. The latter relation means that the beamsplitter effectively cuts all laser phase noise introducing, however, its own displacement noise.

In an experiment homodyne detector $\textrm{D}_C$ measures the quadrature components of $c^{\textrm{r}}_{\textrm{out}}$. However, keeping this in mind, we will deal with the field amplitude itself, since calculations with quadratures result in very cumbersome formulas, while not changing the physical meaning of the ultimate results.

Now let us derive the transmitted signals. Since they are detected by two different devices, each of the signals should be calculated in the proper reference frame of the corresponding detector. Keeping in mind that $\delta x_{\textrm{BS}}$ and $\delta y_{\textrm{BS}}$ should be explicitly specified for each of these reference frame, we substitute Eqs. (\ref{eq_mean_amplitudes} -- \ref{eq_input_wave_vertical}) into formula (\ref{eq_transmitted_signal_FP}) and obtain:
\begin{align*}
    a^{\textrm{t}}_{\textrm{out}}&=
    \frac{1}{\sqrt{2}}\mathcal{T}\Bigl[a_{\textrm{L}}-c_{\textrm{vac}}-
    A_{\textrm{L}0}ik_0\left(\xi_{\textrm{L}}+\xi^{\textrm{f.t.}}_{\textrm{gw}}
    -i\Omega\tau\xi_{\textrm{D}_A}\right)\Bigr]\nonumber\\
    &\quad+R^2Te^{3i\omega_0\tau}A_{+0}2ik_0\,
    \frac{(\xi_{b_1}+\xi^{\textrm{r.t.}}_{\textrm{gw}})e^{i\Omega\tau}-\xi_{a_1}}
    {\mathcal{T}^2_{\omega_0+\Omega}}\,e^{i\Omega\tau}\nonumber\\
    &\quad+A^{\textrm{t}}_{\textrm{out}0}ik_0\xi_{\textrm{D}_A}+
    \mathcal{R}a_{\textrm{vac}},\nonumber\\
    %%%%%%%%%%%%%%%%%%%%%%%%%%%%%%%%%%%%%%%%%%%%%%%%%%%%%%%%%%%%%%%%%%%%%%%%%%%%%%%%%%%%%%%
    b^{\textrm{t}}_{\textrm{out}}&=
    \frac{1}{\sqrt{2}}\mathcal{T}(a_{\textrm{L}}+c_{\textrm{vac}})+
    \mathcal{R}b_{\textrm{vac}}\nonumber\\
    &\quad-\frac{1}{\sqrt{2}}\mathcal{T}A_{\textrm{L}0}ik_0
    (\xi_{\textrm{L}}-\xi_{\textrm{BS}}+\eta_{\textrm{BS}}-\xi^{\textrm{f.t.}}_{\textrm{gw}}
    -i\Omega\tau\eta_{\textrm{D}_B})\nonumber\\
    &\quad+R^2Te^{3i\omega_0\tau}B_{+0}2ik_0\,
    \frac{(\eta_{b_2}-\xi^{\textrm{r.t.}}_{\textrm{gw}})e^{i\Omega\tau}-\eta_{a_2}}
    {\mathcal{T}^2_{\omega_0+\Omega}}\,e^{i\Omega\tau}\nonumber\\
    &\quad+B^{\textrm{t}}_{\textrm{out}0}ik_0\eta_{\textrm{D}_B}.
\end{align*}
Two different regimes of detection of transmitted signals are possible.
\begin{enumerate}
    \item{Resonant regime. Let both cavities be tunes to resonance. In this case amplitude detector measures $A^{\textrm{t}}_{\textrm{out}0}(a^{\textrm{t}}_{\textrm{out}})^\dag+(A^{\textrm{t}}_{\textrm{out}0})^\dag a^{\textrm{t}}_{\textrm{out}}$ coinciding with the amplitude quadrature $\sim a^{\textrm{t}}_{\textrm{out}}+ (a^{\textrm{t}}_{\textrm{out}})^\dag$ measured by the homodyne detector, since $A^{\textrm{t}}_{\textrm{out}0}$ is a pure real quantity. Resonant regime means that we are tuned to the peak of the resonant curve. At this operating point variation of optical wave amplitude is very weak ($\sim\Omega^2$). Therefore, neither amplitude detector nor homodyne detector measuring amplitude quadrature can be used. Instead, one should use the homodyne detector that measures the phase quadrature $\sim a^{\textrm{t}}_{\textrm{out}}-(a^{\textrm{t}}_{\textrm{out}})^\dag$. However, in this case all the homodyne detectors should be synchronized with enough accuracy so that they have equal mean phase. Otherwise, different detectors will measure slightly different quadrature components.}
    \item{Non-resonant regime. In this case we are tuned to the slope of the resonance curve and variation of amplitude of the optical wave is $\sim\Omega$. Amplitude detection can be used then. Its major advantage is that it does not require synchronization between different amplitude detectors. One can also use homodyne detectors to measure amplitude or phase quadratures. It this case, however, synchronization between detectors will be required.}
\end{enumerate}
Let detectors $\textrm{D}_A$ and $\textrm{D}_B$ be homodyne detectors for definiteness so that they measure the quadratures of $a^{\textrm{t}}_{\textrm{out}}$ and $b^{\textrm{t}}_{\textrm{out}}$ correspondingly. The reference oscillations should be produced by some local sources, for instance, lasers that have the same carrier frequency $\omega_0$ and are synchronized with laser L. As usually required for the homodyne detectors, the amplitudes of these local oscillators are assumed to be much larger than the mean output amplitudes $A^{\textrm{t}}_{\textrm{out}0}$ and $B^{\textrm{t}}_{\textrm{out}0}$. In this case we can neglect their intrinsic noises (laser noises). However, they are required to be synchronized, i.e. have the same homodyne phase in order to measure identical quadratures.

Once the quadratures are measured they can be stored in a computer memory and later processed. For instance, an experimentalist may produce any desired linear combinations between them. Let's consider the cosine quadratures of the signals for definiteness. A simple subtraction of $a^{\textrm{t}}_{\textrm{out}}$ quadrature from $b^{\textrm{t}}_{\textrm{out}}$ quadrature, evidently, cancels the term containing laser phase noise. Similar, one can operate with the sine quadratures. This can be thought of as a possible method of laser noise cancelation from transmitted waves. In the case of reflected waves elimination of laser noise takes place at the level of interference of field amplitudes and further recording of laser-noiseless field in the form of quadratures. In the case of transmitted waves we first record the quadratures containing laser noise and then linearly combine them to produce the laser-noise-free quantity. However, the change in a sequence of procedures (to combine first and then record or first record and then combine) does not introduce any meaningful physical difference, therefore, in theory we may operate with transmitted field amplitudes without the need to perform cumbersome calculations with their quadratures.

Keeping this in mind we construct the following combination of the fields:
\begin{align}
    d^{\textrm{t}}_{\textrm{out}}&=\frac{1}{\sqrt{2}}\,b^{\textrm{t}}_{\textrm{out}}-
    \frac{1}{\sqrt{2}}\,a^{\textrm{t}}_{\textrm{out}}\nonumber\\
    &=\mathcal{T}c_{\textrm{vac}}+
    \frac{1}{\sqrt{2}}\,\mathcal{R}(b_{\textrm{vac}}-a_{\textrm{vac}})+
    \mathcal{T}A_{\textrm{L}0}ik_0\xi^{\textrm{f.t.}}_{\textrm{gw}}\nonumber\\
    &\quad-\frac{1}{2}\,\mathcal{T}A_{\textrm{L}0}ik_0
    \bigl[\eta_{\textrm{BS}}-\xi_{\textrm{BS}}-
    i\Omega\tau(\eta_{\textrm{D}_B}-\xi_{\textrm{D}_A})\bigr]\nonumber\\
    &\quad+\frac{1}{\sqrt{2}}\,R^2Te^{3i\omega_0\tau}A_{+0}\,2ik_0\,
    \frac{\eta_{b_2}-\xi_{b_1}-2\xi^{\textrm{r.t.}}_{\textrm{gw}}}
    {\mathcal{T}^2_{\omega_0+\Omega}}\,e^{2i\Omega\tau}\nonumber\\
    &\quad-\frac{1}{\sqrt{2}}\,R^2Te^{3i\omega_0\tau}A_{+0}\,2ik_0\,
    \frac{\eta_{a_2}-\xi_{a_1}}{\mathcal{T}^2_{\omega_0+\Omega}}\,e^{i\Omega\tau}\nonumber\\
    &\quad+A^{\textrm{t}}_{\textrm{out}0}ik_0(\eta_{\textrm{D}_B}-\xi_{\textrm{D}_A}).
    \label{eq_transmitted_signal_Michelson}
\end{align}
Here we assumed that $A_{+0}=B_{+0}$ due to equal characteristics of the cavities. Again, one can establish the equivalence between this differential signal with the transmitted signal (\ref{eq_transmitted_signal_FP}) of a single cavity.

\subsection{Cancelation of displacement noise of the end mirrors in a single interferometer} Now we have two laser-noise-free signals (\ref{eq_reflected_signal_Michelson}) and (\ref{eq_transmitted_signal_Michelson}) which can be combined to cancel one of the fluctuative degrees of freedom. Since there are four such quantities, $\eta_{\textrm{D}_B}-\xi_{\textrm{D}_A}$, $\eta_{b_2}-\xi_{b_1}$, $\eta_{a_2}-\xi_{a_1}$ and $\eta_{\textrm{BS}}-\xi_{\textrm{BS}}$, we should somehow suppress two more degrees of freedom \textit{by hands} (the last one will be eliminated by the additional interferometer, see below). We introduce the following model assumptions (see Fig. \ref{pic_Michelson_FP}):
\begin{enumerate}
    \item{Both the input mirrors are rigidly attached to beamsplitter, i.e. $\xi_{a_1}=\xi_{\textrm{BS}}$ and $\eta_{a_2}=\eta_{\textrm{BS}}$. The composite mass will be called platform $\textrm{P}_{\textrm{BS}}$ with associated fluctuative degree of freedom $\eta_{\textrm{P}_{\textrm{BS}}}-\xi_{\textrm{P}_{\textrm{BS}}}$}.
    \item{Detectors $\textrm{D}_A$ and $\textrm{D}_B$ are rigidly attached to the end-mirrors $b_1$ and $b_2$ respectively, i.e. $\xi_{b_1}=\xi_{\textrm{D}_A}$ and $\eta_{b_2}=\eta_{\textrm{D}_B}$. Corresponding platforms will be called $\textrm{P}_1$ and $\textrm{P}_2$ and their differential degree of freedom $\eta_{\textrm{P}_2}-\xi_{\textrm{P}_1}$.}
\end{enumerate}
The realizability of these requirements remains an open practical question.

Let us now substitute the introduced relations between displacements into signals (\ref{eq_reflected_signal_Michelson}) and (\ref{eq_transmitted_signal_Michelson}) and rewrite them in terms of the input amplitude $A_{\textrm{L}0}$:
\begin{align*}
    c^{\textrm{r}}_{\textrm{out}}&=\mathcal{R}c_{\textrm{vac}}+
    \frac{1}{\sqrt{2}}\,\mathcal{T}(b_{\textrm{vac}}-a_{\textrm{vac}})\nonumber\\
    &\quad-\frac{1}{2}\,\frac{RT^2e^{2i\omega_0\tau}}
    {\mathcal{T}^2_{\omega_0}\mathcal{T}^2_{\omega_0+\Omega}}\,
    A_{\textrm{L}0}2ik_0(\eta_{\textrm{P}_2}-\xi_{\textrm{P}_1})e^{i\Omega\tau}\nonumber\\
    &\quad+\frac{1}{2}\,\frac{RT^2e^{2i\omega_0\tau}}
    {\mathcal{T}^2_{\omega_0}\mathcal{T}^2_{\omega_0+\Omega}}\,
    A_{\textrm{L}0}ik_0(\eta_{\textrm{BS}}-\xi_{\textrm{BS}})(1+e^{2i\Omega\tau})\nonumber\\
    &\quad+\frac{1}{2}\,\frac{RT^2e^{2i\omega_0\tau}}
    {\mathcal{T}^2_{\omega_0}\mathcal{T}^2_{\omega_0+\Omega}}\,
    A_{\textrm{L}0}\frac{\omega_0}{\Omega}\,h(e^{2i\Omega\tau}-1),\nonumber\\
    %%%%%%%%%%%%%%%%%%%%%%%%%%%%%%%%%%%%%%%%%%%%%%%%%%%%%%%%%%%%%%%%%%%%%%%%%%%%%%%%%%%%%%%
    d^{\textrm{r}}_{\textrm{out}}&=\mathcal{T}c_{\textrm{vac}}+
    \frac{1}{\sqrt{2}}\,\mathcal{R}(b_{\textrm{vac}}-a_{\textrm{vac}})\nonumber\\
    &\quad+\frac{1}{2}\,\frac{T^2e^{i\omega_0\tau}}
    {\mathcal{T}^2_{\omega_0}\mathcal{T}^2_{\omega_0+\Omega}}\,
    A_{\textrm{L}0}ik_0(\eta_{\textrm{P}_2}-\xi_{\textrm{P}_1})\nonumber\\
    &\qquad\times\Bigl(1+R^2e^{2i(\omega_0+\Omega)\tau}+
    i\Omega\tau\mathcal{T}^2_{\omega_0}e^{i\Omega\tau}\Bigr)\nonumber\\
    &\quad-\frac{1}{2}\,\frac{T^2e^{i\omega_0\tau}}
    {\mathcal{T}^2_{\omega_0}\mathcal{T}^2_{\omega_0+\Omega}}\,
    A_{\textrm{L}0}ik_0(\eta_{\textrm{BS}}-\xi_{\textrm{BS}})
    (1+R^2e^{2i\omega_0\tau})e^{i\Omega\tau}\nonumber\\
    &\quad-\frac{1}{2}\,\frac{T^2e^{i\omega_0\tau}}
    {\mathcal{T}^2_{\omega_0}\mathcal{T}^2_{\omega_0+\Omega}}\,
    A_{\textrm{L}0}\frac{\omega_0}{\Omega}\,h\nonumber\\
    &\qquad\times\biggl[\mathcal{T}^2_{\omega_0}(e^{i\Omega\tau}-1)+
    R^2e^{2i\omega_0\tau}(e^{2i\Omega\tau}-1)\nonumber\\
    &\qquad\qquad-\frac{1}{2}\,(\Omega\tau)^2\mathcal{T}^2_{\omega_0}e^{i\Omega\tau}\biggr],
\end{align*}

From these signals we can exclude either $\eta_{\textrm{BS}}-\xi_{\textrm{BS}}$ or $\eta_{\textrm{P}_2}-\xi_{\textrm{P}_1}$. The following linear combination cancels the later quantity:
\begin{align}
    s_1&=c^{\textrm{r}}_{\textrm{out}}
    \Bigl(1+R^2e^{2i(\omega_0+\Omega)\tau}+
    i\Omega\tau\mathcal{T}^2_{\omega_0}e^{i\Omega\tau}\Bigr)\nonumber\\
    &\quad+2d^{\textrm{r}}_{\textrm{out}}Re^{i(\omega_0+\Omega)\tau}\nonumber\\
    &=s_1^{\textrm{vac}}+\frac{1}{2}\,\frac{RT^2e^{2i\omega_0\tau}}
    {\mathcal{T}^2_{\omega_0}\mathcal{T}^2_{\omega_0+\Omega}}\,
    A_{\textrm{L}0}ik_0(\eta_{\textrm{BS}}-\xi_{\textrm{BS}})\nonumber\\
    &\qquad\times\biggl[\mathcal{T}^2_{\omega_0+\Omega}(1-e^{2i\Omega\tau})
    +i\Omega\tau\mathcal{T}^2_{\omega_0}(1+e^{2i\Omega\tau})e^{i\Omega\tau}\biggr]\nonumber\\
    &\quad-\frac{1}{2}\,\frac{RT^2e^{2i\omega_0\tau}}
    {\mathcal{T}^2_{\omega_0}\mathcal{T}^2_{\omega_0+\Omega}}\,
    A_{\textrm{L}0}\frac{\omega_0}{\Omega}\,h\nonumber\\
    &\qquad\times\Bigl[\mathcal{T}^2_{\omega_0+\Omega}(1-e^{i\Omega\tau})^2
    +i\Omega\tau\mathcal{T}^2_{\omega_0}(1-e^{2i\Omega\tau})e^{i\Omega\tau}\nonumber\\
    &\qquad\qquad-(\Omega\tau)^2\mathcal{T}^2_{\omega_0}e^{2i\Omega\tau}\Bigr],
    \label{eq_end_mirr_noise_free_signal}
\end{align}
where $s_1^{\textrm{vac}}$ is the combined vacuum noise.

It is straightforward to verify that in the long-wave ($\Omega\tau\ll1$) and narrow-band ($T^2=2\gamma\tau\ll1$, where $\gamma$ is the cavity half-bandwidth) approximations signal $s_1$ reduces to:
\begin{align*}
    s_1&=2c_{\textrm{vac}}+\sqrt{2}(b_{\textrm{vac}}-a_{\textrm{vac}})\nonumber\\
    &\quad-\frac{\gamma}{\gamma-i\delta}\,A_{\textrm{L}0}\,
    \frac{1}{(\gamma-i\delta-i\Omega)\tau}\nonumber\\
    &\quad\times ik_0(\Omega\tau)^2
    \left(\eta_{\textrm{BS}}-\xi_{\textrm{BS}}+\frac{1}{2}\,Lh\right),
\end{align*}
where $\delta$ is detuning from resonance.

\subsection{Cancelation of displacement noise of beamsplitter platform in a double interferometer}\label{subsec_noise_cancel_double} 
\begin{figure}[h]
\begin{center}
\includegraphics[scale=0.54]{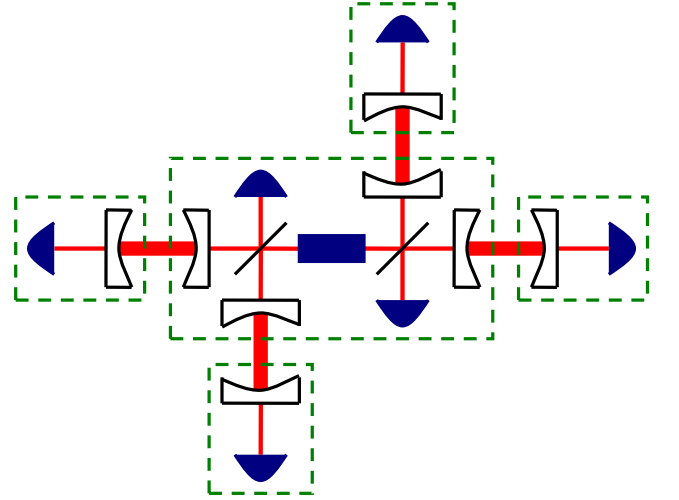}
\caption{A double Michelson/Fabry-Perot interferometer having common
central platform.} \label{pic_double_Michelson_FP}
\end{center}
\end{figure}

The tidal structure of metric (\ref{eq_metric_tensor}) immediately suggests the method of cancelation of beamsplitter platform noise. Consider a scheme with two Michelson/Fabry-Perot interferometers having common central platform (see Fig. \ref{pic_double_Michelson_FP}).
Let us assume that we have eliminated displacement noise of the end-platforms of the second (left-bottom) interferometer and obtained the signal $s_2$ containing the fluctuations of central platform and the GW. This signal can be evaluated straightforwardly from formula (\ref{eq_end_mirr_noise_free_signal}) replacing $(\eta_{\textrm{BS}}-\xi_{\textrm{BS}})\rightarrow-(\eta_{\textrm{BS}}-\xi_{\textrm{BS}})$ and keeping the GW function $h$ unchanged due to the symmetry of GW wavefront. Ultimately, adding $s_2$ to $s_1$ we obtain signal $s$ free from displacement noise of the cental (beamsplitters) platform:
\begin{align}
    s=s^{\textrm{vac}}-&\frac{RT^2e^{2i\omega_0\tau}}
    {\mathcal{T}^2_{\omega_0}\mathcal{T}^2_{\omega_0+\Omega}}\,
    A_{\textrm{L}0}\frac{\omega_0}{\Omega}\,h\nonumber\\
    &\times\Bigl[\mathcal{T}^2_{\omega_0+\Omega}(1-e^{i\Omega\tau})^2
    +i\Omega\tau\mathcal{T}^2_{\omega_0}(1-e^{2i\Omega\tau})e^{i\Omega\tau}\nonumber\\
    &\qquad-(\Omega\tau)^2\mathcal{T}^2_{\omega_0}e^{2i\Omega\tau}\Bigr].
    \label{eq_DFI_MFP_noise_free_signal}
\end{align}
Here $s^{\textrm{vac}}$ describes total vacuum noise in both interferometers. In long-wave and narrow-band approximations we obtain:
\begin{align}
    s&=2\Bigl[c^{(1)}_{\textrm{vac}}+c^{(2)}_{\textrm{vac}}\Bigr]+
    \sqrt{2}\Bigl[b^{(1)}_{\textrm{vac}}+b^{(2)}_{\textrm{vac}}-
    a^{(1)}_{\textrm{vac}}-a^{(2)}_{\textrm{vac}}\Bigr]\nonumber\\
    &\quad-\frac{\gamma}{\gamma-i\delta}\,A_{\textrm{L}0}\,
    \frac{1}{(\gamma-i\delta-i\Omega)\tau}\,ik_0(\Omega\tau)^2\,\frac{1}{2}\,Lh.
    \label{eq_DFI_MFP_noise_free_signal_long-wave}
\end{align}
Here vacuum fields with upper index $(1)$ denote the vacuum fluctuations in detector ports of the first interferometer and vacuum fields with index $(2)$ denote the vacuum fluctuations in the corresponding ports of the second interferometer.

\begin{figure}[!h]
\begin{center}
\includegraphics[scale=0.50]{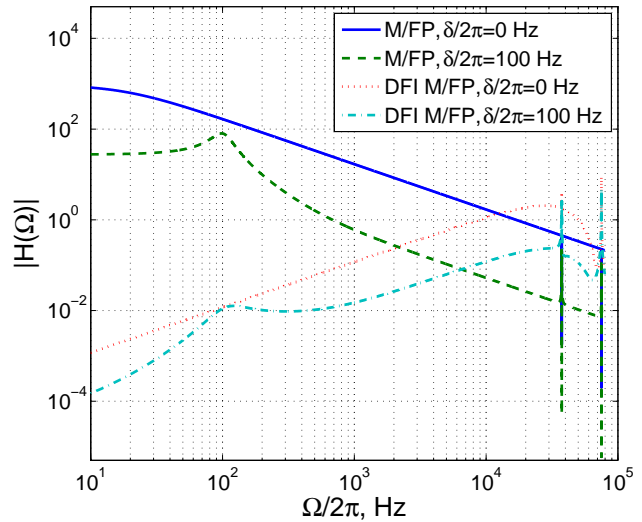}
\caption{Absolute values of GW transfer functions for conventional Michelson/Fabry-Perot interferometer ($\delta/2\pi=0,\ 100$ Hz), and for double Michelson/Fabry-Perot DFI ($\delta/2\pi=0,\ 100$ Hz).}\label{pic_transfer_function}
\end{center}
\end{figure}

It is convenient for methodological purposes to compare the susceptibilities to GWs of the considered interferometer and of the conventional interferometer with Michelson/Fabry-Perot topology (without any recycling mirrors) which response is described by the formula \cite{2007_GW_FP_LL}:
\begin{align}
    s(\omega_0+\Omega)&=-A_{\textrm{L}0}\,
    \frac{T^2e^{2i\delta\tau}}{1-Re^{2i\delta\tau}}\,
    \frac{1}{1-Re^{2i(\Omega+\delta)\tau}}\nonumber\\
    &\qquad\times ik_0Lh(\Omega)\,\frac{\sin\Omega\tau}{\Omega\tau}\,
    e^{i\Omega\tau}.
    \label{eq_MFP_response}
\end{align}
Let the readout schemes in both interferometers register the following quadrature(s):
\begin{equation*}
    \mathfrak{s}(\Omega)=\frac{s(\omega_0+\Omega)-s^\dag(\omega_0-\Omega)}{\sqrt{2}i},
\end{equation*}
where $\mathfrak{s}(\Omega)$ is given either by formula (\ref{eq_DFI_MFP_noise_free_signal}) for displacement-noise-free double Michelson/Fabry-Perot topology (without $s^{\textrm{vac}}$ term), or (\ref{eq_MFP_response}) for conventional Michelson/Fabry-Perot topology. To compare the GW sensitivities we define the transfer function $H(\Omega)$ as the ratio of GW response quadrature $\mathfrak{s}(\Omega)$ to $A_{\textrm{L}0}ik_0Lh(\Omega)$. We plotted the absolute values of both transfer functions in Fig. \ref{pic_transfer_function} for the following parameters (most close to Advanced LIGO): $L=4\times10^3$ km and $R=0.997$. For comparison we chose two values of detuning for each system: $\delta/2\pi=0,\ 100$ Hz.

Fig. \ref{pic_transfer_function} clearly demonstrates strong GW suppression at low frequencies according to the predicted $\Omega^2$-law in Eq. (\ref{eq_DFI_MFP_noise_free_signal_long-wave}). However, at higher frequencies $\sim 10^4$ Hz both traditional and DFI topologies acquire approximately equal level of GW susceptibility.

\section{Discussion}
Let us briefly summarize all the essential model assumptions that we used for our gedanken experiment.

\begin{enumerate}
    \item For a single Michelson/Fabry-Perot interferometer we assumed that the end-photodetectors are rigidly attached to the end-mirrors and input mirrors are rigidly attached to the beamsplitter. For a pair of interferometers we assumed that both beamsplitters and all the input mirrors are mounted of the common central platform. Although these assumptions do not contradict any fundamental principles, the question of their practical realization is highly questionable, at least for the ground-based facilities. In addition, even such a composite platform does not allow full cancelation of its internal thermal noise (our substitution $(\eta_{\textrm{BS}}-\xi_{\textrm{BS}})\rightarrow-(\eta_{\textrm{BS}}-\xi_{\textrm{BS}})$ is valid for displacement of the center of mass only). However, in principle, one may think of constructing a space-based interferometer with arm-cavity lengths of several hundreds meters or kilometers which will be most sensible to GWs at frequencies below 1 Hz. At such low frequencies the model of rigid platform may look more attractive from the experimental point of view than that at higher frequencies in the Earth-bound environment. Another way to soften the complexity of the optical scheme is to ``squeeze'' geometrically the additional interferometer so that both interferometers share the same beamsplitter. Nevertheless, this does not cancel the requirement that all the input mirrors are attached to the beamsplitter.
    \item If the end-detectors of the cavities operate as homodyne detectors then the local oscillators are assumed to be present. The frequency of these local oscillators should coincide with the carrier frequency of the source laser, so they should be kept synchronized with it. In addition all the homodyne detectors themselves should be synchronized with each other in such a way that they have identical homodyne phase, otherwise, they will measure different quadrature components. The use of amplitude detection looks much more attractive from the practical point of view.
\end{enumerate}

\section{Conclusion}
In this paper in a form of gedanken experiment we have analyzed the operation of a double Michelson/Fabry-Perot interferometer performing the laser- and displacement-noise-free gravitational-wave detection. It has been demonstrated that if certain model requirements met (input mirrors and beamsplitters can be rigidly mounted on a single platform and end detectors can be rigidly attached to the end mirrors) it is possible to construct such a linear combination of interferometer responses (their quadratures) that produces the strongest displacement-noise-free response to the gravitational wave allowed by general relativity. Namely, the DFI response function turns out to be proportional to $(f_{\textrm{gw}}\tau)^2/(\gamma\tau)$, where $f_{\textrm{gw}}$ is the GW frequency, $\tau=L/c$ with $L$ being the length of interferometer arms and $\gamma$ is the cavity bandwidth. However, the question of practical realizability of our model assumptions is open for criticism.

\acknowledgements The authors would like to thank F.Ya. Khalili and S.L. Danilishin for valuable critical remarks and comments on the paper. We also would like to express our gratitude to A. Freise, S. Hild and S. Chelkowski for the hospitality and support during our stay at Birmingham University and the inspiring discussions which greatly helped to improve our research.

%\bibliography{noise_free}

\end{document}